# Super-resolution Orthogonal Deterministic Imaging Technique for Terahertz Subwavelength Microscopy

Hichem Guerboukha[1], Yang Cao[1], Kathirvel Nallappan[1,2], and Maksim Skorobogatiy[1,*]

Engineering physics, Polytechnique Montréal, Montréal, Canada

Electrical engineering, Polytechnique Montréal, Montréal, Canada

*maksim.skorobogatiy@polymtl.ca*

**Abstract:** Terahertz subwavelength imaging aims at developing THz microscopes able to resolve deeply subwavelength features. To improve the spatial resolution beyond the diffraction limit, a current trend is to use various subwavelength probes to convert the near-field to the far-field. These techniques, while offering significant gains in spatial resolution, inherently lack the ability to rapidly obtain THz images due to the necessity of slow pixel-by-pixel raster scan and often suffer from low light throughput. In parallel, in the visible spectral range, several super-resolution imaging techniques were developed that enhance the image resolution by recording and statistically correlating multiple images of an object backlit with light from stochastically blinking fluorophores. Inspired by this methodology, we develop a Super-resolution Orthogonal Deterministic Imaging (SODI) technique and apply it in the THz range. Since there are no natural THz fluorophores, we use optimally designed mask sets brought in proximity with the object as artificial fluorophores. Because we directly control the form of the masks, rather than relying on statistical averages, we aim at employing the smallest possible number of frames. After developing the theoretical basis of SODI, we demonstrate the second-order resolution improvement experimentally using phase masks and binary amplitude masks with only 8 frames. We then numerically show how to extend the SODI technique to higher orders to further improve the resolution. As our formulation is based on the equation of linear imaging and it uses spatial modulation of either the phase or the amplitude of the THz wave, our methodology can be readily adapted for the use with existing phase-sensitive single pixel imaging systems or any amplitude sensitive THz imaging arrays.

## 1. Introduction

Terahertz science and technology (0.1-10 THz, wavelengths of 3 mm-30 μm) are now mature research fields with many fundamental and practical applications in sensing and imaging [1,2]. THz subwavelength imaging, in particular, aims at developing THz microscopes able to resolve deeply subwavelength features [3]. To improve the spatial resolution beyond the diffraction limit, a current trend in the THz research is to use various subwavelength probes such as apertures [4], metallic tips [5], solid immersion lenses [6], dielectric cuboids [7], etc. that essentially facilitate the creation and scattering of an evanescent subwavelength-sized near-field probing wave into the far-field. While these techniques offer significant gains in spatial resolution, they inherently lack the ability to rapidly obtain THz images, due to the necessity of slow pixel-by-pixel raster scans, and often long averaging times caused by low measured signal intensities.

To overcome these challenges, several designs of subwavelength imaging systems have been proposed where the sample is in direct contact with a multi-pixel detector for parallel pixel acquisition. In one of these approaches, the sample is brought in contact with a nonlinear crystal which is then used for two-dimensional electro-optic sampling with a sensitive visible CCD camera [8, 9]. Recently, a fully integrated THz near-field camera was developed where both the emitter and detector were fabricated on the same chip using SiGe heterojunction bipolar transistor technology [10]. While promising, these techniques require a direct physical contact between the sample and the detector, which is practically inconvenient.

Computational imaging techniques have also been developed for applications in the terahertz range, including single-pixel imaging and compressive sensing [11]. There, a semiconductor substrate is used as a spatial light modulator when optically pumped [12]. Specially designed patterns obtained using a digital micromirror device are then used as illumination masks for imaging with single-pixel detectors. Moreover, subwavelength imaging can be achieved by placing the sample in direct contact with the mask, allowing to encode near-field features in the reconstructed image [13, 14]

In parallel, in the visible spectral range, several practically important super-resolution imaging techniques have been developed to beat the diffraction limit [15, 16]. These techniques use temporally uncorrelated blinking fluorophores distributed spatially as subwavelength probes, and they involve a computational treatment of a collection of images (frames) that represent snapshots of the target object backlit by randomly blinking fluorophores. One of such techniques is known as super-resolution optical fluctuation imaging (SOFI) [17, 18]. By assuming that the fluorescent labels switch rapidly and stochastically between binary on/off states, a higher-resolution image is reconstructed from a collection of frames, using the statistics of cumulants. The key advantage of this technique is that one forgoes the slow raster scanning with subwavelength probes or image acquisition in the object near-field, while only collecting intensity images in the far-field. This allows realizing super-resolution imaging using only far-field acquisitions with amplitude-sensitive detector arrays (that are now also available in the THz range [19, 20]), which is both practical and convenient in most industrial settings.

For these reasons, we explore in this paper the possibility of adopting the SOFI technique to the THz spectral range. An immediate problem that we face is the lack of natural fluorophores in the THz range, with the exception of some exotic atomic optical fluorescence [21]. Nonetheless, since the THz wavelengths are relatively large, we propose to use optimal ensembles of "artificial blinking fluorophores" in the form of judiciously designed amplitude or phase masks brought in close proximity with the object. While in this paper, we fabricate both amplitude and phase masks using etching and 3D printing techniques, amplitude masks can also be realized dynamically via spatial light modulation as used in THz single-pixel imaging and compressive sensing [13]. Importantly, since we directly control the form of the masks, rather than relying on statistical averages of a large number of frames, as it is done in SOFI, we rather aim at employing the smallest possible number of optimally designed super-resolution masks to reduce the number of frames necessary to deterministically reconstruct the image. This is important, because limited frame acquisition rate is one of the key challenges of the SOFI technique even in the visible spectral range.

We also note that the proposed technique which we refer to as Super-resolution Orthogonal Deterministic Imaging (SODI), while superficially similar to compressive sensing, is different from it conceptually and experimentally. Thus, in compressive imaging one generates a number of random masks that is typically a fraction of the desired number of pixels, leading to loss of information during the image reconstruction. A single-pixel detector can be used within compressive sensing technique. In contrast, within SODI, one uses the smallest possible number of subwavelength-structured "orthogonal" masks necessary for deterministic image reconstruction up to any desired resolution order. Within SODI, higher order of the reconstruction means subwavelength resolution proportional to the wavelength divided by the square root of the order number, while the super-resolved imaging is reconstructed without loss of information. Additionally, within SODI, amplitude sensitive (pyroelectric, bolometer, etc.) or phase sensitive (EOS/CCD) detector arrays can be used in the far-field.

On the other hand, SODI employs an ensemble of masks with subwavelength features and in this respect, it bears resemblance to techniques that employ near-field to far-field conversion using subwavelength probes. That said, individual masks in SODI show high light throughput (100% phase masks and 50% amplitude masks), which is significantly higher compared to other standard subwavelength probes such as apertures and needles. Additionally, raster scanning of the object with a subwavelength probe is replaced by consecutive image acquisitions in the far-field using an ensemble of optimally designed masks that cover the whole image.

The paper is organized as follows. First, in Section 2, we present theoretical foundation behind the SODI technique, and detail the design of optimal ensembles of amplitude and phase masks that mimic sets of stochastically blinking fluorophores. Then, in Section 3, we present an example of phase masks and demonstrate experimental super-resolution imaging at the second order. Next, in Section 4, we show how to modify the super-resolution algorithm to be able to use binary amplitude (on/off) masks, and again demonstrate super-resolution imaging at the second order. Finally, in Section 5, we demonstrate how to improve the resolution further, by providing algorithms for amplitude and phase mask design for higher order super-resolution imaging.

## 2. Super-resolution reconstruction using deterministic fluctuations, second order formulation

The following presentation is largely based on the mathematical foundation behind the SOFI technique [17, 18]. Considering a linear imaging system, the measured image $E(\vec{r})$ is the spatial convolution of the object $O(\vec{r})$ with the system's impulse response $S(\vec{r})$, also known as the point spread function (PSF):

$$E(\vec{r}) = \int d\vec{r}_1 S(\vec{r} - \vec{r}_1) \cdot O(\vec{r}_1) M(\vec{r}_1, t) \qquad (1)$$

Within the SOFI technique $M(\vec{r}, t)$ in Eq. 1 corresponds to the time-dependent spatial distribution of light intensity emitted by deeply subwavelength stochastically blinking fluorophores. In our SODI adaptation of this technique, $M(\vec{r}, t)$ corresponds to the complex transmission function of a given subwavelength-structured mask labeled with index $t \in [1, N_t]$ where $N_t$ is the number of masks in a measurement set (also equal to the number of images to be acquired during the experiment). We start by considering the variance of the image. By denoting $\langle ... \rangle_t$ to be a simple average over a collection of frames, we obtain:

$$
\begin{aligned}
\langle E^2(\vec{r}, t) \rangle_t &- \langle E(\vec{r}, t) \rangle_t^2 \\
&= \iint d\vec{r}_1 d\vec{r}_2 S(\vec{r} - \vec{r}_1) S(\vec{r} - \vec{r}_2) O(\vec{r}_1) O(\vec{r}_2) \\
&\cdot (\langle M(\vec{r}_1, t) M(\vec{r}_2, t) \rangle_t - \langle M(\vec{r}_1, t) \rangle_t \langle M(\vec{r}_2, t) \rangle_t)
\end{aligned}
\qquad (2)
$$

In general, taking the square of the image does not result in a higher-resolution image. It is a somewhat subtle point as, for example, a squared grayscale images will appear to have sharper/narrower boundaries, however, if two lines in such an image appear as one, the same will be observed in the squared image. To get higher resolution within the SOFI technique one assumes that any two fluorophores blink independently. Then, one can define the following averaging function over a time period $\Delta t$ and the corresponding orthogonality relation between light intensities emitted by the two point-size fluorophores:

$$
\begin{aligned}
\langle M(\vec{r}_1, t) M(\vec{r}_2, t) \rangle_t &- \langle M(\vec{r}_1, t) \rangle_t \langle M(\vec{r}_2, t) \rangle_t \\
&= \frac{1}{\Delta t} \int_0^{\Delta t} dt \, M(\vec{r}_1, t) M(\vec{r}_2, t) - \left( \frac{1}{\Delta t} \int_0^{\Delta t} dt \, M(\vec{r}_1, t) \right) \left( \frac{1}{\Delta t} \int_0^{\Delta t} dt \, M(\vec{r}_2, t) \right) \\
&= C \cdot \delta(\vec{r}_1 - \vec{r}_2)
\end{aligned}
\qquad (3)
$$

where $C$ is assumed to be a spatially independent constant and $\delta(\vec{r})$ is the Delta function. With this orthogonality relation, the reconstructed image [Eq. (2)] becomes:

$$
\begin{aligned}
\langle E^2(\vec{r}, t) \rangle_t &- \langle E(\vec{r}, t) \rangle_t^2 \\
&= \frac{1}{\Delta t} \int_0^{\Delta t} dt \, E^2(\vec{r}, t) - \left( \frac{1}{\Delta t} \int_0^{\Delta t} dt \, E(\vec{r}, t) \right)^2 \\
&= C \cdot \int d\vec{r}_1 S^2(\vec{r} - \vec{r}_1) O^2(\vec{r}_1)
\end{aligned}
\qquad (4)
$$

Since a PSF squared is narrower than the original PSF, one will observe improvement in the image resolution. For example, frequently, the PSF can be expressed as a Gaussian of width $\sigma$:

$$S(\vec{r}) = \exp\left( -\frac{|\vec{r}|^2}{2\sigma^2} \right) \qquad (5)$$

Therefore, taking the square of that Gaussian yields another Gaussian of a reduced width $\tilde{\sigma} = \sigma/\sqrt{2}$, thereby leading to an improved resolution.

Similarly, within SODI, by judiciously designing a set of mutually orthogonal masks and by properly choosing the frame averaging operator $\langle ... \rangle_t$, we can obtain a higher-resolution image reconstruction as will be demonstrated in the rest of this section. In the following, we detail several design principles for the construction of the optimal mask sets $M(\vec{r}, t)$ for SODI technique. First, we assume that the PSF has a bounded support:

$$S(\vec{r} - \vec{r}_1) S(\vec{r} - \vec{r}_2) = 0 \quad \text{if} \quad |\vec{r}_1 - \vec{r}_2| > D \qquad (6)$$

where $D$ is the characteristic width of the PSF, which is typically set by diffraction on the imaging optics, and, thus, $D \propto \lambda$. This also means that two features will be automatically resolved if they are well separated in space by at least a distance $D$ or further (see the example of two Gaussian-like PSFs in Fig. 1a). In the case of the SOFI technique, it also means that to achieve resolution enhancement, one must require that the emission from closely placed point-size fluorophores within a distance $D$ or closer to each other are uncorrelated. At the same time, fluorophores that are positioned further than $D$ from each other will not affect resolution even if they blink in a correlated manner. In the case of the SODI technique, this means that the mask orthogonality should only be forced locally, within any spatial domain of characteristic size $D$, while, the same mask set does not have to be orthogonal over longer distances:

$$\langle M(\vec{r}_1, t)M(\vec{r}_2, t)\rangle_t = \begin{cases} 0 & \text{if} \quad |\vec{r}_1 - \vec{r}_2| < D \\ C & \text{if} \quad \vec{r}_1 = \vec{r}_2 \end{cases} \tag{7}$$

This realization leads to a significant reduction in the number of required masks/frames to enable resolution enhancement. Thus, a locally orthogonal mask set can be constructed using a basic pixel group, and then be periodically repeated to cover the whole object under imaging. According to Eq. (7), the physical size of the basic pixel group must be comparable to the PSF size. Furthermore, we are not limited to using square tiling of the basic pixel groups, as triangular or hexagonal tiling can also be used to cover the two-dimensional plane (Fig. 1b).

Note that the orthogonality definition given by Eq. (7) is somewhat different from that used in Eq. (3). As we will see in the following, Eq. (7) is more appropriate for pure phase masks, while Eq. (3) will be used later in the paper for amplitude masks.

Next, we consider digital masks that feature a small number of pixels in the basic pixel group. This means that pixel positions in the mask can accept a finite and discrete number of values $N_r$, and we label individual pixels with an index $r$ (see Fig. 1b). Then, the local orthogonality condition for the digital masks (assuming periodic tiling of the basic pixel group to cover the whole plane) can be, for example, expressed as:

$$\langle M(\vec{r}_1, t)M(\vec{r}_2, t)\rangle_t = \frac{1}{N_t}\sum_{t=1}^{N_t} M(r_1, t)M(r_2, t) = \begin{cases} C & \text{if } r_1 = r_2 \\ 0 & \text{if } r_1 \neq r_2 \end{cases} \quad ; \quad r_1, r_2 \in [1, N_r] \tag{8}$$

where $N_t$ is the total number of masks, $t$ is the mask number, while $r_1, r_2$ are the pixel numbers within the basic pixel group, and $C$ is a common constant that is independent of pixel position. The mask orthogonality condition in the form of Eq. (8) is mostly suitable for pure phase masks. Using a mask set that satisfies Eq. (8), one has to acquire $N_t$ far-field images $E(\vec{r}, t)$ of the object covered by the different masks in this set. Then, an image of the object with increased resolution can be reconstructed using:

$$\langle E^2(\vec{r}, t)\rangle_t = \frac{1}{N_t}\sum_{t=1}^{N_t} E^2(\vec{r}, t) = C \cdot \int d\vec{r}_1 S^2(\vec{r} - \vec{r}_1)O^2(\vec{r}_1) \tag{9}$$

As in the SOFI technique, the resolution is improved because the PSF is squared in Eq. (9). The mask set can be better visualized using a matrix notation, where the matrix rows correspond to the individual masks, the columns correspond to the individual pixels in the basic pixel group, while the matrix values are the corresponding $M(r, t)$. For example, a locally orthogonal set containing $N_t = 2^k$ pure phase masks can be constructed using Hadamard matrix (Fig. 1c), where the columns are all mutually orthogonal vectors in the sense of Eq. (8). At the same time, we are not forced to use all the columns of the Hadamard matrix to form a locally orthogonal set of masks, so in fact the number of pixels $N_r$ in the basic pixel group can be smaller than the number of masks $N_t$. At the same time, using periodic tiling of the basic pixel set ensures that no matter what pixel is chosen in the object plane, a thus designed mask set will always be locally orthogonal in the sense of Eq. (8). As an example, in Fig. 1b we show three different choices of basic pixel groups that contain $N_r = 4,6,7$ pixels that can be tiled periodically to fill the 2D plane. Then using Hadamard matrix (Fig. 1c), one can construct locally orthogonal sets of $N_t = 8$ phase masks (with either 0 or $\pi$ phase shifts) containing $N_r$ pixels in the corresponding basic pixel group (Fig. 1d). By tiling the basic pixel groups periodically to fill the whole 2D plane, one arrives at a complete set of locally orthogonal masks shown in Fig. 1e. Finally, in order to ensure that all pixels in a basic pixel set contribute equally to the image reconstruction we require that the constant $C$ in Eq. (8) is the same for all the pixels in a given mask set. In other words, this ensures that on average the object is uniformly illuminated by the masks.

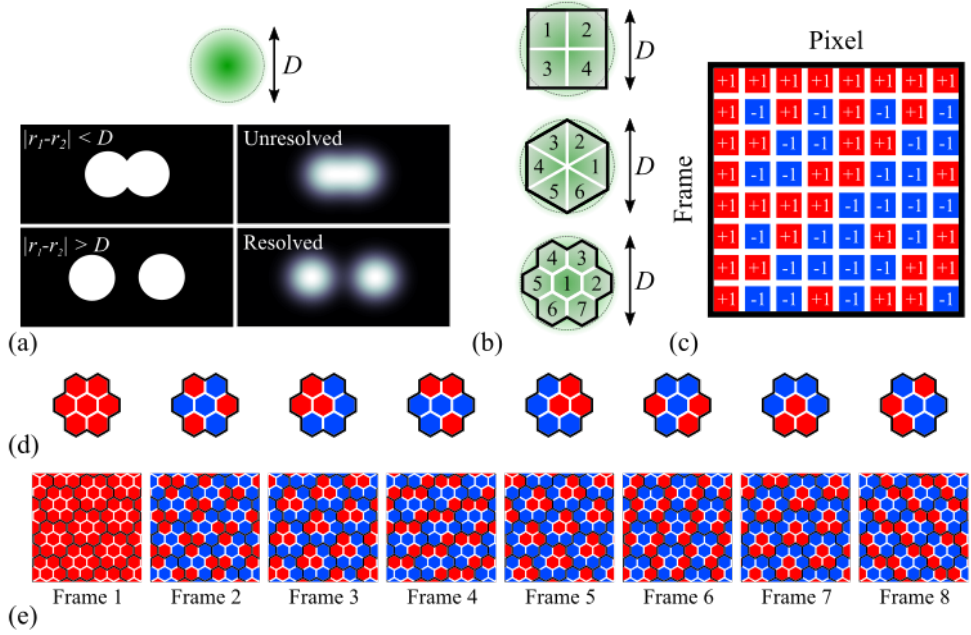

**Fig. 1** Super-resolution image reconstruction with deterministic fluctuations. (a) Two features are resolved if they are separated in space further away than the characteristic size $D$ of the PSF spread function. (b) Examples of basic pixel groups containing 4 square pixels, 6 triangular pixels, and 7 hexagonal pixels for periodic tiling of the 2D plane. (c) Hadamard matrix used in the second-order reconstruction. Here, different matrix rows define pixel values in the basic pixel groups of different masks. (d) Basic pixel groups of the 8 masks based on the Hadamard basis, and (e) corresponding 8 complete masks where the basic pixel groups are periodically tiled to cover the whole field-of-view.

## 3. Pure phase masks

In Fig. 2 we present experimental results of the SODI image reconstruction measured using a THz time-domain imaging system (see Supplementary section A for more details about the experimental setup). We used Hadamard basis $(0, \pi)$ phase masks as described in the previous section. As an object we use a cutout in metal in the form of the fleur-de-lys (see Supplementary section B). As explained above, the 8 frames are formed using 7 pixels derived from the Hadamard basis arranged in a hexagonal lattice with an inter-pixel distance of 2 mm (Fig. 2a). The basic pixel group covers a circle of size ~6mm, which is comparable to the PSF size for our imaging system as measured using the knife-edge technique (see Supplementary section C for details). To realize –1 element of the Hadamard basis, we use phase elements fabricated by 3D printing hexagonal steps of height $h$ using PLA plastic of refractive index $n \approx 1.6$. The incurred phase due to passage through a step of thickness $h$ is $\phi = 2\pi(n-1)h/\lambda$, which indicates that to obtain a phase difference of $\pi$ at 0.32 THz ($\lambda \approx 0.94$ mm), we must use $h = 800$ μm. In Fig. S2b of the Supplementary material B, we show 8 phase masks that correspond to 8 rows of the Hadamard basis/matrix.

We then use THz time-domain spectroscopy for both amplitude and phase super-resolution image reconstruction. In particular, for every mask in the 8 mask set, we place the given phase mask on top of the object and then image the pair using raster scanning in the focal plane between two parabolic mirrors (see Supplementary material A and Fig S1 for more details). In Fig. 2b, we show amplitude images of the resulting interference patterns, while in Fig. 2c the corresponding phase distributions are presented. Next, following Eq. (9), we compute the super-resolved image by taking the average of the squares of the complex fields from each measurement $\langle E^2(r, t) \rangle$. We note that we average the complex fields squared and not the intensities, which allows us to retrieve super-resolved images of both amplitude and phase distributions. The original image amplitude and phase are shown in Fig. 2d, while our SODI images are shown in Fig. 2e,f where we present both $\langle E^2(\vec{r}, t) \rangle$, and $\sqrt{\langle E^2(\vec{r}, t) \rangle}$ distributions. We note that $\sqrt{\langle E^2(\vec{r},t)\rangle}$ preserve the same intensity contrast as the original image, while providing better resolution than the original image. Finally, to better see the effect of our algorithm on the resolution, we plot the amplitude and phase cross-sections along a line in the image where the three leaves meet. In the original image, the three leaves can hardly be distinguished, while in the super-resolved amplitude and phase images they are clearly resolvable.

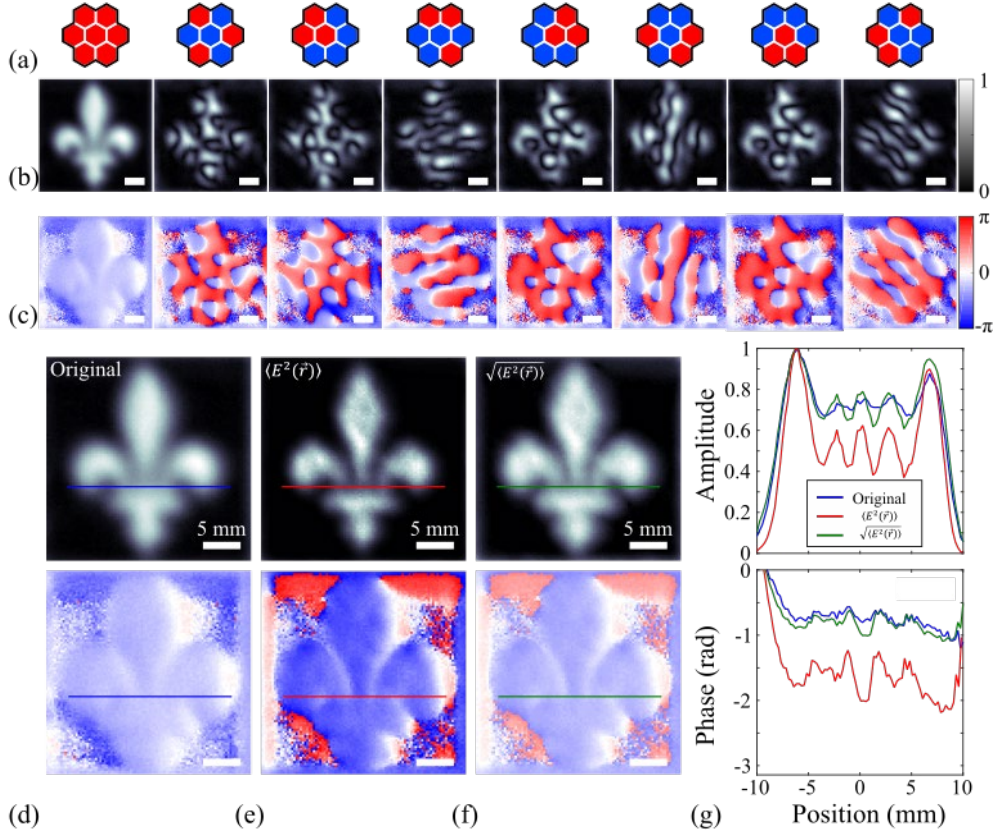

**Fig. 2** Second order image reconstruction using phase masks based on Hadamard basis. (a) Basic pixel groups of each mask. (b) Amplitude and (c) phase of the corresponding measured frames. (d) Amplitude (top) and phase (bottom) of the original image, (e) the super-resolution reconstruction using $\langle E^2 \rangle$ and (f) $\sqrt{\langle E^2 \rangle}$ images. (g) Linear section of the amplitude (top) and phase (bottom) distributions along the line where the three leaves meet. Scale bar size is 5 mm.

## 4. Binary amplitude masks

In the previous section, we detailed second order SODI technique using a set of pure phase masks based on Hadamard orthogonal basis. In this section, we detail an alternate second-order SODI reconstruction that uses binary amplitude masks that are also related to the Hadamard basis. The case of binary amplitude masks is important as it can be efficiently realized using THz amplitude spatial light modulators based on dynamic photomodulated masks [12-14].

We now demonstrate how starting from the Hadamard basis that contains +1 and -1 elements (Fig. 1c), we can modify the SODI technique and derive an orthogonal set of amplitude masks that contains only 0 and 1 elements and features ~50% light transmission through each mask. The 0 and 1 elements can be interpreted as opaque and transparent regions of a binary amplitude mask, and they can be realized experimentally as cutouts in an otherwise non-transparent metal foil. Particularly, one can show that by simply replacing the -1 elements by 0 in the Hadamard basis of size $N_t$, and after omitting the first column of 1's, we arrive at a set of $N_t$ binary amplitude masks $M(\vec{r}_1, t)$ each containing $N_r = N_t - 1$ pixels that are locally orthogonal in the following sense:

$$\langle M(\vec{r_1},t)M(\vec{r_2},t)\rangle_t - \langle M(\vec{r_1},t)\rangle_t\langle M(\vec{r_2},t)\rangle_t$$

$$= \frac{1}{N_t}\sum_{t=1}^{N_t} M(r_1,t)M(r_2,t) - \frac{1}{N_t}\sum_{t=1}^{N_t} M(r_1,t)\frac{1}{N_t}\sum_{t=1}^{N_t} M(r_2,t) \qquad (10)$$

$$= \begin{cases} C & \text{if } r_1 = r_2 \\ 0 & \text{if } r_1 \neq r_2 \end{cases} \quad ; \quad r_1, r_2 \in [1, N_r]$$

where the constant $C = 3/4$ for the binary amplitude mask set derived from the Hadamard basis of any order $N_t$.

The SODI technique modified for amplitude mask sets then requires a subtraction of the squared mean of the measured electric field from the mean of the square of the electric field, where averaging is performed over the set of $N_t$ amplitude masks:

$$\langle E^2(\vec{r},t)\rangle_t - \langle E(\vec{r},t)\rangle_t^2$$

$$= \iint d\vec{r_1}d\vec{r_2}S(\vec{r}-\vec{r_1})S(\vec{r}-\vec{r_2})O(\vec{r_1})O(\vec{r_2})[\langle M(\vec{r_1},t)M(\vec{r_2},t)\rangle - \langle M(\vec{r_1},t)\rangle\langle M(\vec{r_2},t)\rangle] \qquad (11)$$

$$= C \cdot \iint d\vec{r_1}S^2(\vec{r}-\vec{r_1})O^2(\vec{r_1})$$

In Fig. 3, we present experimental results of the SODI reconstruction using binary amplitude masks derived from the Hadamard basis (Fig. 3a). Measured at 0.32 THz amplitude and phase distributions for a fleur-de-lys image superimposed with different masks are shown in Fig.3b and Fig. 3c respectively. The amplitude of the original image and its corresponding super-resolution reconstruction are shown in Fig. 3d at various frequencies from 0.27 THz to 0.42 THz. The SODI images for the reconstructed phases are shown in Supplementary material D. Unlike the phase masks which are designed to operate at a specific THz frequency (since the phase of the underlying phase element is frequency dependent), the SODI reconstruction using amplitude mask sets can be performed over a larger bandwidth. However, when increasing the THz frequency, we start observing artifacts in the reconstruction image in the form of multiple dots (high spatial frequency noise). These appear because the PSF size at higher frequencies becomes smaller than the size of the basic pixel group of the patterned mask, therefore the local orthogonality condition between blinking patterns of the individual pixels Eq. (10) might not hold anymore due to partial overlap of the PSF spatial support with the individual pixels.

Finally, although in our experiments we used a THz-TDS imaging system, and were, therefore, capable of super-resolution reconstruction of both amplitude and phase distributions, the SODI technique is not restricted to coherent measurements and it can be performed using field intensities instead of electric fields. Indeed, our super-resolution algorithm does not specify the nature of the frames $E(\vec{r},t)$. As long as the image can be represented as the convolution of some measurand with a given PSF [Eq. (1)], we can apply our super-resolution algorithm. For example, in Fig. 3e, we show SODI reconstruction using the maximum amplitude value of the peak THz electric field in time domain. This measurand is often used to reconstruct THz images as it provides rapid images without the need to move the delay line of the THz-TDS system [14]. Again, we observe the resolution improvement using only the peak THz electric field.

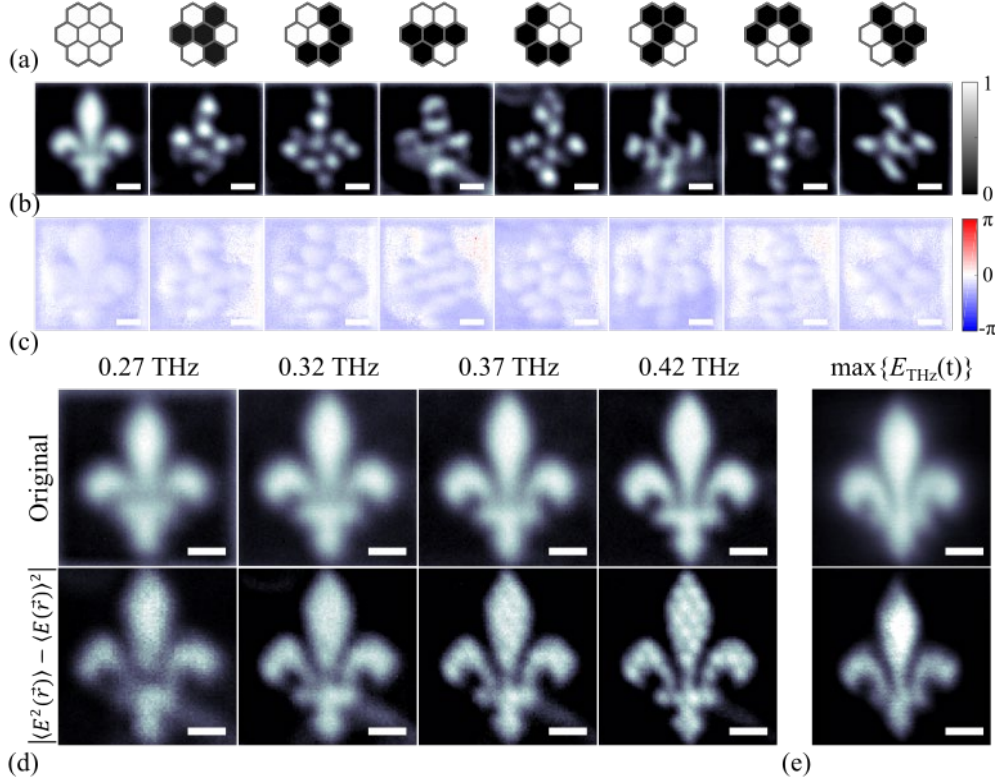

**Fig. 3** Second order image reconstruction using binary amplitude masks constructed using a modified Hadamard basis. (a) Basic pixel groups in each mask. (b) Amplitude and (c) phase of the corresponding measured frames. (d) Comparison of the amplitude of the original images (top) and their corresponding super-resolution reconstructions using complex electric field (bottom) at different frequencies and (e) when considering only the maximum electric field peak value in time domain. The super-resolution reconstructions of the phase are shown in Supplementary section D. Scale bar size is 5 mm.

## 5. Super-resolution reconstruction using deterministic fluctuations, higher order formulation

In the previous sections, we have detailed the second order SODI algorithm for image resolution enhancement. Naturally, one can construct higher order super-resolution schemes by using higher moments in the definition of the algorithm. In the following we present one possible formulation that is most applicable to the case of phase masks. Particularly, for the $n^{\text{th}}$-order algorithm, one needs to calculate:

$$
\begin{aligned}
&\langle E^n(\vec{r},t)\rangle_t \\
&= \iint d\vec{r}_1 d\vec{r}_2 \dots \vec{r}_n \\
&\cdot S(\vec{r}-\vec{r}_1)S(\vec{r}-\vec{r}_2)\dots S(\vec{r}-\vec{r}_n)\, O(\vec{r}_1)O(\vec{r}_2)\dots O(\vec{r}_n)\langle M(\vec{r}_1,t)M(\vec{r}_2,t)\dots M(r_n,t)\rangle_t
\end{aligned}
\tag{12}
$$

The local orthogonality condition for a given mask set can be defined as:

$$
\begin{aligned}
&\langle M(\vec{r}_1,t)M(\vec{r}_2,t)\dots M(\vec{r}_n,t)\rangle_t \\
&= \frac{1}{N_t}\sum_{t=1}^{N_t} M(\vec{r}_1,t)M(\vec{r}_2,t)\dots M(\vec{r}_n,t) \\
&= \begin{cases} C & \text{if } \vec{r}_1 = \vec{r}_2 = \cdots \vec{r}_n \\ 0 & \text{else} \end{cases}
\end{aligned}
\tag{13}
$$

where $M(\vec{r},t)$ is considered a vector of size $N_t$ with elements that describe changing optical properties over time of a fixed pixel/fluorophore located at position $\vec{r}$. Then, the reconstructed image is convoluted with the $n^{\text{th}}$ power of the PSF:

$$\langle E^n(\vec{r}, t) \rangle_t = \iint d\vec{r}_1 S^n(\vec{r} - \vec{r}_1) O^n(\vec{r}_1) \tag{14}$$

As previously, considering a Gaussian PSF of width $\sigma$ [Eq. (5)], taking it to the $n^{\text{th}}$ power results in another Gaussian with a reduced width $\tilde{\sigma} = \sigma/\sqrt{n}$, thus leading to an improved resolution.

Finding a basis that respects the orthogonality condition Eq. (14) for a general order $n$ is not trivial. An obvious choice for any order $n$ is the identity matrix, the columns of which form "blinking" patterns $M(\vec{r}, t)$ of artificial fluorophores. This choice, however, results in binary amplitude masks with very low transmission efficiency as only one pixel in the basic pixel group is 1 (transmitting), while all the other pixels are 0 (blocking the light). This resulting mask set corresponds to a subwavelength aperture successively scanning the object pixel by pixel. Moreover, following Bethe's study of diffraction by a circular hole [22, 23], the transmitted electric field decreases with the third power of the aperture size, thus significantly decreasing the overall signal-to-noise ratio of the measurements. In contrast, light transmission through pure phase masks is significantly more efficient as was demonstrated in Section 3 when using, for example, a Hadamard basis. When increasing the order of the SODI method, one is interested in using mask sets with high light transmission. Additionally, for convenience of the mask fabrication, we seek to use either pure phase or binary amplitude masks.

It is important to mention that the mask orthogonality condition defined by Eq. (13) is not the traditional definition of an orthogonal basis set which requires that the scalar product between any two distinct vectors in a set is zero. In fact, for the SODI method of order $n$, an orthogonal basis set in the sense of Eq. (13) requires that the convolution of any $n$ vectors in a set, among which at least two are different to be equal to zero. To the best of our knowledge, the problem of finding orthogonal basis for higher order orthogonality definition in this sense is still an open problem that does not have a general solution. Finally, we note that higher order super-resolution reconstruction formulations that are alternative to Eq. (12) can be formulated as one only needs a certain general orthogonality condition between the individual blinking patterns of spatially distinct pixels, while the main advantage of Eq. (12) is its simplicity.

In what follows, we present some particular solutions that we found for the problem of high-order orthogonal basis sets in the sense of Eq. (13). We first present a solution for the $3^{\text{rd}}$ order orthogonal basis set containing any desired number of $N_r$ vectors (number of pixels in the basis pixel group). An example of $4^{\text{th}}$-order orthogonal basis set can be found in Supplemental material E. As discussed earlier, temporal changes in the optical properties of a fluorophore (artificial blinker) located at position $\vec{r}$ is described by the function $M(\vec{r}, t)$. By limiting ourselves to a finite group of $N_r$ pixels in the basic pixel group, and considering only $N_t$ temporal snapshots (number of masks), we adopt a matrix notation where $M$ is a rectangular matrix of dimensions $N_t \times N_r$, where the rows correspond to the different time frames, while the columns correspond to the different pixels. Consider now $n$ columns of the matrix $M$ with column indices $r_1, r_2, \dots, r_n$. The orthogonality relation of the $n^{\text{th}}$ order [Eq. (13)] can then be written as:

$$\begin{aligned} &\langle M(r_1, t) M(r_2, t) \dots M(r_n, t) \rangle_t \\ &= \frac{1}{N_t} \sum_{t=1}^{N_t} M(r_1, t) M(r_2, t) \dots M(r_n, t) \\ &= \begin{cases} C & \text{if } r_1 = r_2 = \dots = r_n \\ 0 & \text{else} \end{cases} \quad ; \quad r_1, r_2, \dots, r_n \in [1, N_r] \end{aligned} \tag{15}$$

To verify the $n^{\text{th}}$ order orthogonality condition of Eq. (15) for given matrix of size $N_t \times N_r$, one should compute the $N_r^n$ possible combinations of vector multiplications to ensure that it is always equal to 0, except when all the indices are equal. In fact, due to the commutative property of the product defined in Eq. (15) the number of distinct multiplications is significantly smaller and can be shown to be equal to $\frac{(n + N_r - 1)!}{n!(N_r - 1)!} \sim \frac{N_r^n}{n!}$.

For the $3^{\text{rd}}$ order orthogonality condition, we found a pure phase basis set for any number of pixels $N_r$. The matrix with columns made of basis vectors is constructed by concatenation of the three following matrices. By denoting $I_N$ to be the identity matrix of size $N$ and by denoting $O_{N \times M}$ to be the matrix of size $N$ by $M$ with every element equal to one, the three matrices can be written as:

$$A_1 = O_{N_r \times N_r} - (1 - i)I_{N_r}$$
$$A_2 = O_{N_r \times N_r} - (1 + i)I_{N_r} \qquad (16)$$
$$A_3 = -O_{2(N_r - 3) \times N_r}$$

The matrix of size $(4N_r - 6) \times N_r$ resulting from the concatenation of these three matrices along the temporal dimension forms a $3^{\text{rd}}$ order orthogonal basis in the sense of [Eq. 15]. Compared to the Hadamard basis, this solution contains four different phase elements: $\{+1, -1, +i, -i\}$. The number of required frames is a function of the number of pixels: $N_t = 4N_r - 6$. However, one can see that all the pixels in the last $2(N_r - 3)$ frames are constituted of $-1$ elements. Therefore, one can in principle measure it only once and reuse it $2(N_r - 3)$ times when taking the average. The smallest number of required frames/masks would then be $N_t = 2N_r + 1$.

We show in Fig. 4a, an example of such a basis made of $N_r = 4$ vectors (number of pixels in the basic pixel group) each containing $N_t = 10$ elements (number of masks in the mask set). To demonstrate that this basis set respects the $3^{\text{rd}}$ order orthogonality condition, we compute the 20 distinct combinations of vectors entering the definition [Eq. (15)] with $n = 3$. As we can see in Fig. 4b, the 20 possible multiplications always yield 0, except in the 4 cases when all the three vectors are identical. We then arrange the 4 pixels that make the basic pixel group into a 2x2 square tile and repeat it periodically, which results in 10 distinct masks shown in Fig. 4c.

In Fig. 4e-m, we numerically compare different orders of reconstruction of a 45x45 mm object in the form of a snowflake cutout (Fig. 4d). The convolution of the object with a Gaussian PSF of width $\sigma = 1$ mm (as defined in Eq. (7)) is shown in Fig. 4e. To apply the SODI algorithm, we consider the basic pixel group of $N_r = 16$ pixels arranged in a 4x4 square tile with the diagonal equal to $6\sigma$ which comprises 99.7% of the Gaussian PSF. The square tiles are then periodically patterned to cover the whole object. For a given order of the SODI reconstruction algorithm, we then choose an appropriate orthogonal basis set and the corresponding set of masks. We then compute the convolution of the masks with the object and the Gaussian PSF using Eq. (1). Finally, to obtain higher-resolution images, we compute the $n^{\text{th}}$ order SODI reconstruction using [Eq. (14)]. The result for order 2 is obtained with a Hadamard basis of $N_t = N_r = 16$ phase masks [Fig. 4f], while the result for order 3 is constructed using the matrix described by Eq. (16) with $N_t = (2N_r + 1) = 33$ phase masks [Fig. 4g]. To clearly demonstrate the necessity of using the orthogonal basis sets for super-resolution image reconstruction, in Fig. 4h, we present a $3^{\text{rd}}$ order reconstruction using random phase matrices instead of the orthogonal ones, clearly the resultant reconstruction is incorrect. For order 4 in Fig. 4i, we use the $4^{\text{th}}$ order orthogonal phase-only basis set that we detail in the Supplementary material E with 256 phase masks. Finally, the results for orders 5 to 8 are obtained using a basis set in the form of an identity matrix that corresponds to successive scanning of the object with a subwavelength aperture of size equal to the size of the single pixel (16 amplitude masks).

As one can see, the resolution of the snowflake improves when increasing the order $n$. However, there is a limit in the improvement as seen in Fig. 4j-m. Artifacts in the reconstruction in the form of discrete dots start to appear when the size of the single pixel in the basic pixel group becomes larger than the width of the $n^{\text{th}}$-power PSF $\sigma/\sqrt{n}$. To avoid this problem, one could reduce the size of the individual pixels used in the mask definition at the expense of increasing the number of masks/frames necessary for image reconstruction.

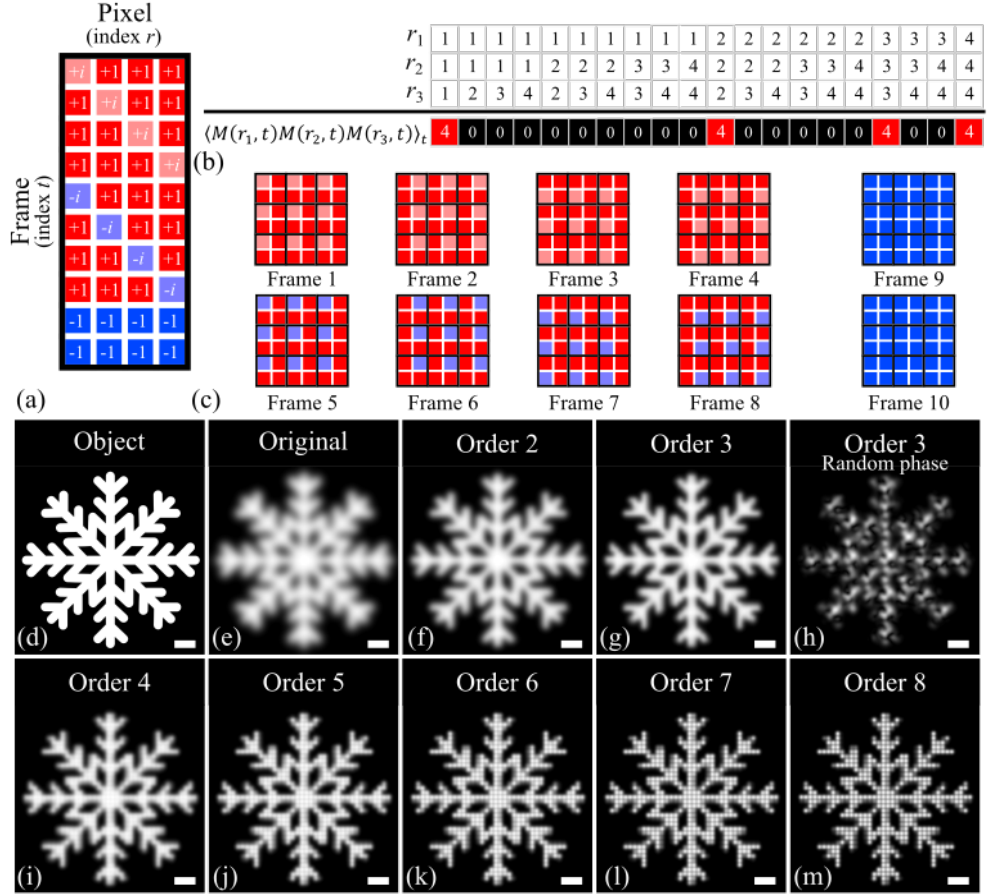

**Fig. 4** Higher order image reconstruction. (a) 3rd order orthogonal basis set (see Eq. (16)) with $N_r = 4$ vectors (number of pixels in the basic pixel group). (b) The third order orthogonality relation $\langle M(r_1,t)M(r_2,t)M(r_3,t)\rangle_t = 0$ is observed when at least one of the vectors is different from the others. (c) 10 locally orthogonal phase masks that correspond to the 3rd order orthogonal basis resulting from periodic pattering of the plane with 2x2 square tiles. (d) Object representing a snowflake cutout, and (e) original image obtained by convolution with a Gaussian PSF of width $\sigma = 1$ mm. (f-n) Super-resolution reconstruction of different orders: (f) Order 2 with Hadamard basis (16 phase masks). (g) Order 3 following Eq. (16) (33 phase masks). (h) Order 3 with random phase masks to show the importance of using locally orthogonal mask sets (33 masks). (i) Order 4 with the pure phase basis detailed in Supplementary material E (256 phase masks). (j) Order 5, (k) Order 6, (l) Order 7, (m) Order 8 with an identity matrix as a basis set (16 binary amplitude masks in the form of a subwavelength aperture). Scale bar size is 5 mm.

## 6. Conclusion

In this paper, we detailed a novel Super-resolution Orthogonal Deterministic Imaging (SODI) technique and demonstrated its application in the THz spectral range. This computational imaging technique is inspired by the Super-resolution Optical Fluctuation Imaging (SOFI) that utilizes stochastic fluctuations in the intensity of the fluorophores integrated into the image under study. However, as there are no natural fluorophores in the THz spectral range, we substitute them by artificial blinkers in the form of judiciously and optimally constructed phase or binary amplitude mask sets. Starting from the equation of linear imaging systems, we find that it is possible to construct super-resolution images using highly transmissive (over 40% throughput) locally orthogonal mask sets. We experimentally demonstrated pure phase mask sets and binary amplitude mask sets resulting in a second order super-resolution reconstruction by processing only 8 images. We also showed how to extend the SODI algorithm to higher orders by using the concept of higher order mask orthogonality. We then numerically demonstrated solutions for the third and fourth order reconstructions using highly transmissive phase masks.

We believe that this work opens new possibilities in THz subwavelength imaging. Using equation for linear imaging systems [Eq. (1)] together with spatial modulation of the phase or amplitude of the THz wave, our methodology can already be applied in different imaging scenarios. For example, it can be directly translated to single-pixel imaging by modifying the photomodulated illumination masks [12-14]. Moreover, as we demonstrated experimentally, this technique can be used with incoherent measurements, which opens up new possibilities for subwavelength imaging using commercially available THz thermal cameras and THz field-effect transistor-based cameras. Finally, this work can be of interest to a larger optical community interested in other wavelengths, as it provides a clear theoretical framework for structuring illumination at a subwavelength scale using optimal phase and amplitude mask sets in order to improve image resolution to any order beyond the diffraction limit.

December 6, 2019

# Supplementary information

## A. Terahertz imaging system

The amplitude and phase images were obtained using a standard THz time-domain spectroscopy (THz-TDS) imaging system [Fig. S1]. An ultrafast Ti:Sapphire laser (Spectra-physics, 800 nm, 100 MHz, 100fs) delivered 300 mW of optical power to an interdigitated photoconductive antenna (BATOP) for THz emission, and 10 mW to a wrapped dipole antenna (Menlo) for detection. The THz was focused on the object and the modulating mask under imaging using a pair of parabolic mirrors. The object was placed on a movable 3D micro-positioning stage and raster-scanned for imaging. An optical delay line was placed on the emitter side and the detected photocurrent was proportional to the electric field of the THz pulse. A Fourier transform of this electric field provided amplitude and phase spectra over a large bandwidth.

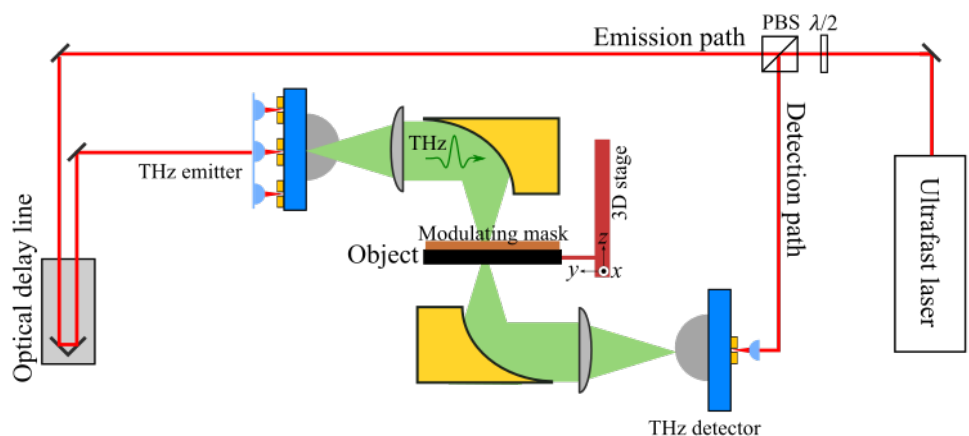

**Fig. S1** THz time-domain spectroscopy system used to obtain spectral amplitude and phase images.

## B. Fabrication of the object and the masks

In our experiments, the binary metal object representing the fleur-de-lys was made using toner-assisted metal foil transfer, also known as hot stamping [Fig. S2a]. This technique allows to fabricate metallic features directly on a paper substrate using a metal foil composite deposited on a thin layer of thermoplastic. The design was first printed on paper with a conventional office laser printer. The metallic foil composite (Therm O Web Deco Foil) was then placed on top of the print and both passed through a laminator. Due to enhanced local heating, the toner and thermoplastic are bonded, leading to a direct imprint of the metal layer on top of the printed design.

The phase masks were fabricated using fused deposition modeling (3D printing) of a polylactic acid (PLA) with refractive index of ~1.6. The phase elements $+1$ and $-1$ were obtained by varying the thickness from 1000 μm to 1800 μm, corresponding to 0 and $\pi$ phase variations at 0.32 THz [Fig. S2b]. The amplitude masks were fabricated using the hot stamping technique discussed above [Fig. S2c].

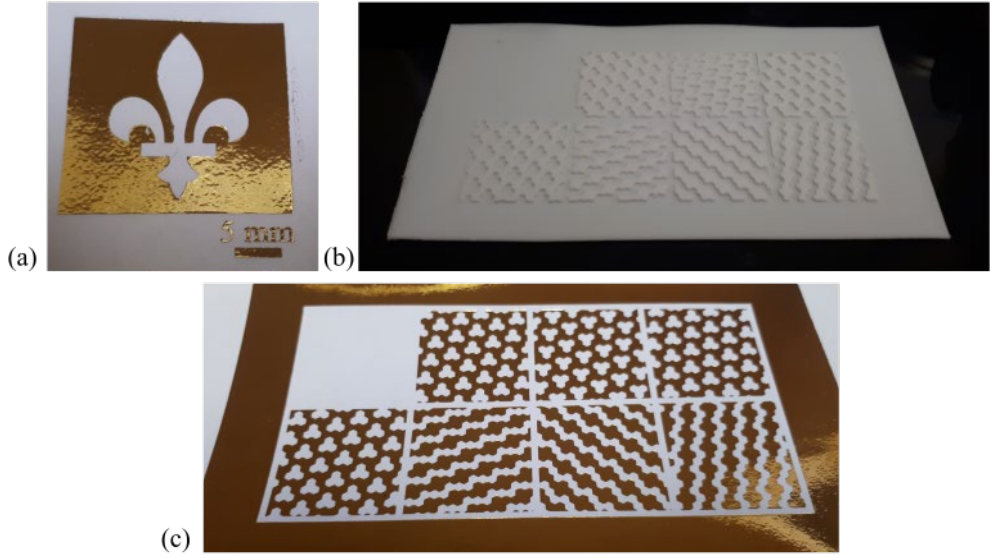

**Fig. S2** (a) Fleur-de-lys object obtained using the hot stamping technique. The bar size is 5 mm. (b) A set of eight pure phase masks for second order SODI fabricated from PLA using fused deposition modeling. (c) A set of eight binary amplitude masks for second order SODI fabricated using toner-assisted metal foil transfer.

## C. Determination of the size of the group of pixels in relation to the point spread function

In the SODI algorithm, the local orthogonality condition is used for the basic pixel group that has a characteristic size comparable to that of the imaging system point spread function (PSF). The basic pixel group is then periodically patterned to cover the whole object.

To determine the size of the basic pixel group for mask construction, we first need to measure the size of the imaging system PSF. For that, we performed a knife-edge measurement of the focused THz beam. We positioned a metal object (knife) in the focal plane and we moved it along the $x$ direction while recording the THz field amplitude as a function of frequency [Fig. S3a]. In the $x$ direction, we assume that the electric field amplitude can be represented with a Gaussian function with a standard deviation $\sigma$:

$$S(x) = A \exp\left[-\frac{(x-x_0)^2}{2\sigma^2}\right] \tag{S1}$$

where $A$ is a normalization parameter and $x_0$ accounts for the center position of the Gaussian. The part of the beam that is measured (that is not blocked by the knife) is then given by the cumulative distribution function of the Gaussian function [24]:

$$\Phi(x) = A \int_{-\infty}^{x} \exp\left[-\frac{(x'-x_0)^2}{2\sigma^2}\right] dx' = \frac{A}{2}\left[1 + \text{erf}\left(\frac{x-x_0}{\sigma\sqrt{2}}\right)\right] \tag{S2}$$

where $x'$ can be seen as the position of the knife in the focal plane. The function $\text{erf}(\xi)$ is the error function defined as:

$$\text{erf}(\xi) = \frac{2}{\sqrt{\pi}} \int_0^{\xi} e^{-t^2} dt \tag{S3}$$

Using Eq. (S2) for every frequency, we fit the measurements using the error function [Eq. (S2)] and retrieve the $\sigma$ parameter of the corresponding Gaussian distribution. We show in Fig. S3b the obtained $\sigma$ parameter as a function of the wavelength (black dots). The red line is the theoretical estimate of the diffraction-limited Gaussian beam waist size [25]:

$$\sigma(\lambda) = \frac{\sqrt{2}}{\pi} \frac{F_m}{D_m} \lambda \tag{S4}$$

where in our case $F_m = 101.6$ mm is the focal length of the parabolic mirror and $D_m = 50.8$ mm is its diameter. The good fit between our experimental data and Eq. (S4) indicates that our system is limited by diffraction.

In the experiments, the basic group of pixels was made of 7 hexagons with an inter-pixel distance of 2 mm. To better visualize how the Gaussian PSF is related in size to the basic group of pixels, we show in Fig. S3(c) the amplitude of the Gaussian PSF [Eq. (S1)] for the different frequencies considered in the experiments, where $\sigma$ was obtained from the measurements of Fig. S3(b). The bold white circle and the dashed white circle correspond to circles of radius $2\sigma$ and $3\sigma$ respectively. Within these circles, 95% ($2\sigma$) and 99.7% ($3\sigma$) of the Gaussian beam amplitude is located. As we can see, for the frequency of 0.32 THz, the basic group of pixels is inscribed inside the circle of $3\sigma$ radius.

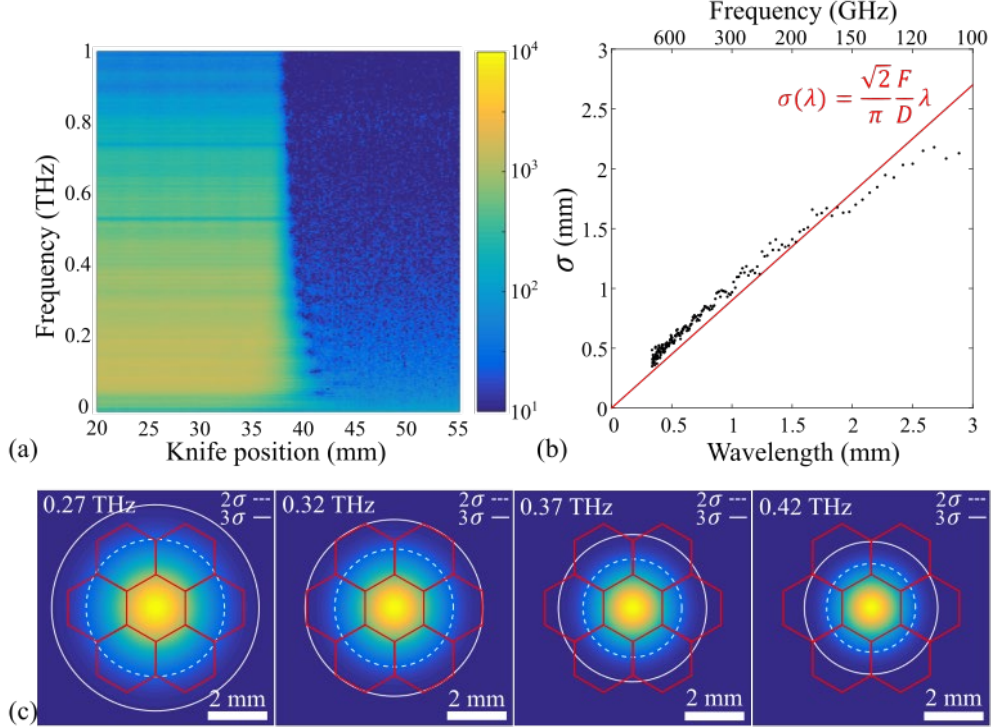

**Fig. S3** (a) Knife-edge measurements. Electric field spectral amplitude as a function of the position of the knife in the focal plane. (b) Standard deviation $\sigma$ as a function of the wavelength. The good fit between the measurements (black dots) and the fit (red line, $\sigma(\lambda) = \frac{\sqrt{2}}{\pi} \frac{F}{D} \lambda$) indicates that the beam is diffraction-limited according to Gaussian beam theory. (d) The basic pixel group used in our experiments and the Gaussian beam distribution for different frequencies.

### D. Reconstruction of the phase for binary amplitude masks

In Section 4 of the main paper, we showed second order SODI reconstruction using binary amplitude masks. In particular, Fig. 3d showed the SODI reconstruction of the amplitude. In this section, we present the phase reconstruction using the same data. The original phase images for different frequencies are shown in the top row of Fig. S4, while the SODI reconstructions when considering the phase of $\langle E^2(\vec{r}) \rangle - \langle E(\vec{r}) \rangle^2$ are presented in the middle row. In all cases, the SODI reconstruction can better resolve the object. To retrieve the original values of the phase while increasing the resolution, one can also compute the phase of $\sqrt{\langle E^2(\vec{r}) \rangle - \langle E(\vec{r}) \rangle^2}$, which is shown in the bottom row of Fig. S4.

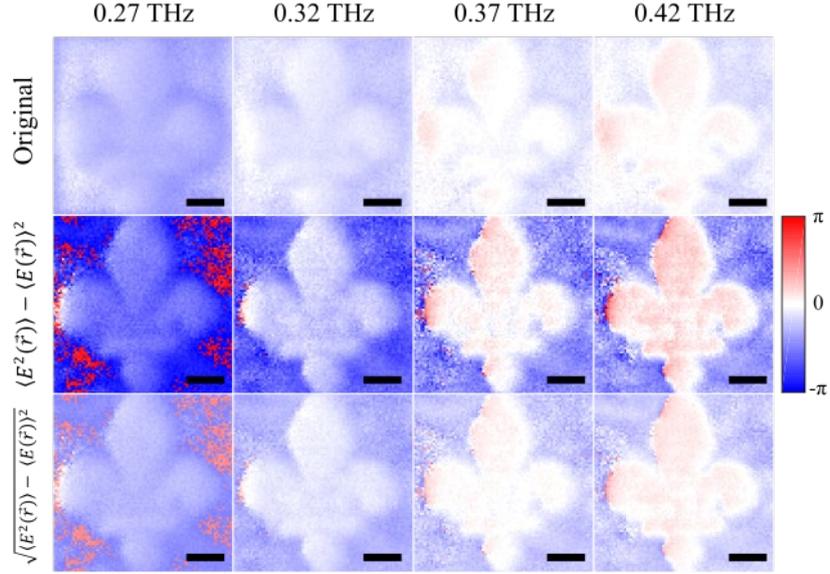

**Fig. S4** Second order SODI phase reconstruction using binary amplitude masks at different frequencies. Top row: original phase images. Middle row: SODI reconstruction using $\langle E^2(\vec{r}) \rangle - \langle E(\vec{r}) \rangle^2$. Bottom row: SODI reconstruction using $\sqrt{\langle E^2(\vec{r}) \rangle - \langle E(\vec{r}) \rangle^2}$. Scale bar size is 5 mm.

## E. Super-resolution reconstruction of order 4

For a reconstruction of order $n = 4$, the $4^{\text{th}}$ order orthogonality condition is defined as:

$$\langle M(r_1, t)M(r_2, t)M(r_3, t)M(r_4, t) \rangle_t = \frac{1}{N_t} \sum_{t=1}^{N_t} M(r_1, t)M(r_2, t)M(r_3, t)M(r_4, t)$$

$$= \begin{cases} C & \text{if } r_1 = r_2 = r_3 = r_4 \\ 0 & \text{else} \end{cases} \tag{S5}$$

For the simulation result shown in Fig. 4i, we constructed a basis with $N_r = 16$ pixels. First, we note that the following basis with two pixels respect the $4^{\text{th}}$ order orthogonality condition:

$$A_1 = \begin{bmatrix} -1 & -1 \\ 1 & -i \\ 1 & -1 \\ i & -1 \end{bmatrix} \tag{S6}$$

as does the following one:

$$A_2 = \begin{bmatrix} -i & -i \\ -i & -1 \\ -1 & -i \\ -1 & 1 \end{bmatrix} \tag{S7}$$

One can then verify that the Kronecker product of these two matrices leads to a new matrix that respects the $4^{\text{th}}$ order orthogonality condition as well:

$$A_3 = A_1 \otimes A_2 = \begin{bmatrix} i & i & i & i \\ i & 1 & i & 1 \\ 1 & i & 1 & i \\ 1 & -1 & 1 & -1 \\ -i & -i & -1 & -1 \\ -i & -1 & -1 & i \\ -1 & -i & i & -1 \\ -1 & 1 & i & -i \\ -i & -i & i & i \\ -i & -1 & i & 1 \\ -1 & -i & 1 & i \\ -1 & 1 & 1 & -1 \\ -1 & 1 & 1 & -1 \\ 1 & -i & i & 1 \\ -i & 1 & 1 & i \\ -i & i & 1 & -1 \end{bmatrix} \tag{S8}$$

This new matrix defines $N_t = 16$ masks containing $N_r = 4$ pixels in the basic pixel group. By continuing in a similar fashion, we can create a fourth matrix by computing the following Kronecker product:

$$A_4 = A_3 \otimes A_1 \tag{S9}$$

where $A_4$ has dimensions of $N_t = 64$ and $N_r = 8$ pixels. Finally, the matrix:

$$A_5 = A_4 \otimes A_2 \tag{S10}$$

has dimensions of $N_t = 256$ (number of locally orthogonal masks in a set) and $N_r = 16$ (number of pixels in the basic pixel group). We verified numerically that the new constructed matrix respects the 4th orthogonality conditions by computing all the 3876 possible distinct 4-vector products. As expected, all of them equal 0, except those when the 4 indices are the same.

Following several numerical experimentations, it seems that the Kronecker product of two matrices that respect the orthogonality condition yields a third matrix that also respect the orthogonality as well. This mathematical hypothesis has yet to be rigorously demonstrated.